\documentclass[aps,pre,twocolumn,showpacs,floatfix,groupedaddress]{revtex4}  % for review and submission 
\usepackage{graphicx}  % needed for figures
\usepackage{dcolumn}   % needed for some tables
\usepackage{bm}        % for math
\usepackage{amssymb}   % for math
\usepackage[latin1]{inputenc}
\usepackage{amsfonts}
\usepackage{amsmath}
\usepackage{amsthm}
\usepackage{color}
\usepackage[american,english]{babel}
\usepackage{fontenc}
\usepackage{textcomp}
\begin{document}

% The following information is for internal review, please remove them for submission
%\widetext
%\leftline{Version .02 as of \today} 
%\leftline{Primary authors: Joe E. Physics}
%\leftline{To be submitted to (PRL, PRD-RC, PRD, PLB; choose one.)}
%\leftline{Comment to {\tt d0-run2eb-nnn@fnal.gov} by xxx, yyy}
%\centerline{\em D\O\ INTERNAL DOCUMENT -- NOT FOR PUBLIC DISTRIBUTION}

% the following line is for submission, including submission to the arXiv!!
%\hspace{5.2in} \mbox{Fermilab-Pub-04/xxx-E}

\title{Continuum percolation of overlapping discs with a  distribution of radii having a  power-law tail}
\author{V. Sasidevan}
\email{sasi@theory.tifr.res.in}
\affiliation{Department of Theoretical 
Physics, Tata Institute of Fundamental Research, Homi Bhabha Road, Mumbai-400005, India.}
%\author{Deepak Dhar}
%\email{ddhar@theory.tifr.res.in}
%\affiliation{Department of Theoretical 
%Physics, Tata Institute of Fundamental Research, Homi Bhabha Road, Mumbai-400005, India.}

%\input list_of_authors_r2.tex  % input Dzero author list
\date{\today}

\begin{abstract}
We study continuum percolation problem of overlapping discs with a  distribution of radii having  a power-law tail; the probability that a given disc has a radius between $R$ and $R+dR$ is proportional to  $R^{-(a+1)}$, where $a > 2$. We show that in the low-density non-percolating phase, the two-point  function shows a power law decay with distance, even at arbitrarily low densities of the discs, unlike the exponential decay in the usual percolation problem. As in the problem of fluids with long-range interaction, we argue that in our problem, the critical exponents take their short range values for $a > 3 - \eta_{sr}$ whereas they depend on $a$ for $a < 3-\eta_{sr}$  where $\eta_{sr}$ is the anomalous dimension for the usual percolation problem. The mean-field regime obtained in the fluid problem  corresponds to the fully covered regime, $a \leq 2$, in the percolation problem. We  propose an approximate renormalization scheme to determine the correlation length exponent $\nu$ and the percolation threshold.
We carry out Monte-Carlo  simulations  and determine the exponent $\nu$ as a function of $a$. The determined values of $\nu$ show that it is independent of the parameter $a$ for $a>3 - \eta_{sr}$ and is equal to that for the lattice percolation problem, whereas $\nu$  varies with $a$ for $2<a<3 - \eta_{sr}$. We also determine the percolation threshold of the system as a function of the parameter $a$. 
\end{abstract}

\pacs{64.60.ah, 64.60.De, 02.50.Ey, 05.10.Ln, 05.70.Fh}
\maketitle 

%\section{\label{sec:level1}First-level heading}
% sections are not used for PRL papers

\section{Introduction}
\label{sec1}

In problems like effective modeling of random media, the continuum models of percolation are more realistic than their lattice counterparts. So, much effort has been put into the study of such systems in the recent past. In two dimensions, the  model systems studied  involve discs, squares etc. of the same size or of different sizes \cite{gawlinski, quintanilla1, roy, phani, balram, ogata} and in three dimensions spheres, cubes etc.,  distributed randomly in space \cite{akagawa, Yi, consiglio, rintoul1, ambrosetti}. An interesting sub-class of problems is where the basic percolating units  have an unbounded size distribution. These are comparatively  less studied, though a few formal results are available \cite{gouere}. The problem of disc percolation where discs have bounded sizes has been studied a lot, mainly by simulation \cite{quintanilla1, quintanilla2,rintoul2}. For the single sized disc percolation, threshold is known to a very high degree of accuracy \cite{quintanilla2}. 
Also simulation studies have shown that the disc percolation in 2D with discs of bounded size falls in the same universality class as that of lattice percolation in 2D \cite{vicsek}. For a review of continuum percolation see \cite{meester1}.  

In this paper we consider  continuum percolation model of overlapping discs in 2D where distribution of the radii of the discs has a power-law tail. We address questions like whether the power-law tail in the distribution of radii changes the critical behavior of the system, and how does the percolation threshold depend on the power of the power-law tail.  From an application point of view, a power-law polydispersity for an extended range of object sizes is quite common in nature especially for fractal systems \cite{martin}. Disordered systems like carbonate rocks often contain pores of widely varied sizes covering many decades in length scales \cite{biswal,roth}, whose geometry may be well modeled by a power-law distribution of pore sizes. The power-law distribution of the radii makes our system  similar to the Ising or fluid system with long-range interactions. For the latter case, it is known that the long-range nature of the interaction does affect the critical behavior of the system for slow enough 
decay of the interaction \cite{aizenman}.  For  similar results in the context of long-range epidemic processes, see \cite{linder}.

The behavior of our model differs from that of the standard continuum percolation model in two aspects. Firstly the entire low density regime in our model shows a power-law decay of the two-point function in contrast to the exponential decay in the standard continuum percolation. Thus the whole low density regime is \textquoteleft critical\textquoteright. However, there is a non-zero percolation threshold below which there is no infinite cluster exist in the system. Secondly the critical exponents are functions of the power $a$ of the power-law distribution for low enough $a$. So while the system belong to the same universality class as the standard continuum percolation for high enough $a$, the critical behavior is quite different for low values of $a$.

The plan of this paper is as follows: In section \ref{sec2}, we define the model of disc percolation  precisely. In section \ref{sec3}, using a rigorous lower bound on the two-point correlation function, we show it decays only as  a power-law with distance for arbitrarily low coverage densities. We discuss the two-point function and critical exponents. In section \ref{sec5}, we propose an approximate renormalization scheme to calculate the correlation length exponent $\nu$ and the percolation threshold in such models. In section \ref{sec6}, we discuss results from simulation  and section \ref{sec7} contains some concluding remarks.

\section{Definition of the model}
\label{sec2}

We consider a  continuum percolation model of overlapping discs in two dimensions. The number density of discs is $n$, and the probability that any small area element $dA$  has  the center of a disc in it is $n dA$, independent of all other area elements. For each disc, we assign  a radius, independently of other discs, from a probability distribution $Prob(R)$. 
We consider the case when $Prob(R)$ has a power-law tail; the probability of radius being greater than $R$ varies as $R^{-a}$ for large $R$. For simplicity, we consider the case when radii take only 
discrete values $R_0 \Lambda^{j}$ where $j = 0,1,2,...$, with probabilities $(1-p) p^j $ where $p = \Lambda^{-a}$. Here $R_0$ is  the size of smallest disc, and $\Lambda$ is a  constant $ >1$.  We call the disc of size $R_0 \Lambda^{j}$  as the disc of type $j$.  %Fig 0 shows  typical realizations of the system for two different number densities with $a = 3$ and $\Lambda = 2.0$.
%\begin{figure}[t]
%\begin{center}
%\includegraphics[scale=0.35]{fig_0.eps} 
%\caption{Two typical realizations of the system of size $L = 200$ with $a = 3$, $\Lambda = 2$ and  number densities $.075$ (on left) and $.20$ (on right). We can easily see that left one is non-percolating and the right one is percolating.}
%\label{fig0}
%\end{center}
%\end{figure}
 
The  fraction of the entire plane which is covered by at least one disc, called the covered area fraction $f_{covered}$,  is given by 
\begin{equation}
f_{covered} = 1 - \exp\left(-A\right)
\label{eq1}
\end{equation}
where $A$ is the areal density - mean area of the discs per unit area of the plane - of the discs, which is finite only for $a > 2$. For $a \leq 2$, in the thermodynamic limit all points of the plane are eventually covered, and $f_{covered} =1$. If $a > 2$, we have areal density,
\begin{equation}
A = n \pi R_0^2 ( 1 - p) / ( 1 - p \Lambda^2)
\label{eq2}
\end{equation}

We define the percolation probability ${\rm P}_{\infty}$ as the probability that a randomly chosen disc belongs to an infinite cluster of overlapping discs. One expects that there is a critical number density $n^*$ such that for $n < n^*$, ${\rm P}_{\infty}$ is exactly zero, but ${\rm P}_{\infty} > 0$, for $ n > n^*$. We shall call the phase $n < n^*$ the non-percolating phase, and the phase $n > n^*$ as the percolating phase.  

It is easy to show that $n^* < \infty$.  We note that for percolation of discs where all discs have the same size $R_0$, there is a  finite critical number density $n_1^*$, such that for $n > n_1^{*}$, ${\rm P}_{\infty} >0$. Then, for the polydisperse case, where all discs have radii $R_0$ or larger, the percolation probability can only increase, and hence $n^* < n_1^*$. Also it has been proved that when ever we have a bounded distribution of radii of the discs, the critical areal density is greater than that for a system with single sized discs \cite{meester2}. Our simulation results show that this remains valid for unbounded distribution of radii of the discs.

\section{The non-percolating phase}
\label{sec3}

We define  two point function ${\rm Prob}(1\rightsquigarrow 2)$ as the probability that points $P_1$ and $P_2$ in the plane  are connected by overlapping discs. Then, by rotational invariance of the problem, ${\rm Prob}(1 \rightsquigarrow 2)$ is only a function of the euclidean distance $r_{12}$ between the two points.  Let ${\rm Prob}^{(1)}( 1 \rightsquigarrow 2)$ denote the probability that there is at least one disc that covers both  $P_1$ and $P_2$. Then, clearly, 
\begin{equation}
{\rm Prob}(1\rightsquigarrow 2) \geq {\rm Prob}^{(1)}( 1 \rightsquigarrow 2).
\label{eq3}
\end{equation}
\begin{figure}
\begin{center}
\includegraphics[scale=0.35]{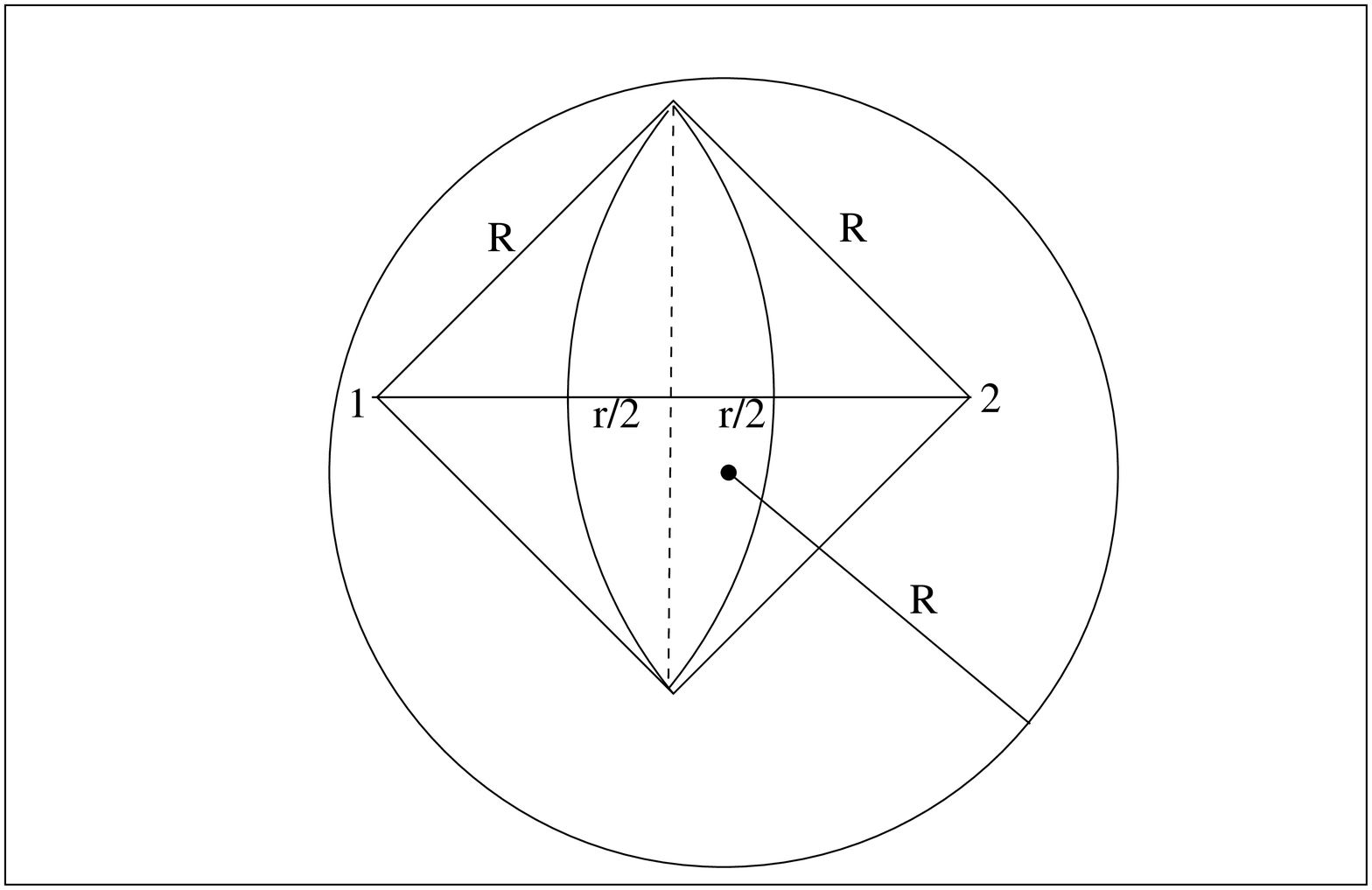} 
\caption{Points 1 and 2 in the plane at a distance $r$ from each other will be covered by a single disc of radius $R$, if the center of such a disc falls in the area of intersection of two circles with radius $R$ and centers at 1 and 2.}
\label{fig1}
\end{center}
\end{figure}
It is straightforward to estimate ${\rm Prob}^{(1)}( 1 \rightsquigarrow 2) $ for our model.  Let $j$ be  the minimum  number  such that radius of disc of type $j$ is greater than or equal to $r_{12}$, i.e. $R_0 \Lambda^{j} \geq r_{12}$. Let $S$ be the region of plane such that the distance of  any point in $S$ from $P_1$ or $P_2$ is less than or equal to $R_{0} \Lambda^{j}$. This region  $S$ is greater than or equal to the  region where each point is within  a distance $r_{12}$ from both $P_1$ and $P_2$.  Using elementary geometry,  the area of region $S$ is greater than or equal to $ (2 \pi/3 -\sqrt{3}/4) r_{12}^2$ (See Fig. \ref{fig1}).  The number density of discs with radius greater than or equal to $R_0 \Lambda^{j}$ is $ n \Lambda^{- a j}$. Therefore, the probability that there is at least one such disc in the region $S$ is $ 1 - \exp\left( -n |S| \Lambda^{-a j}\right)$, where $|S|$ is the area of region $S$.
Thus we get,
\begin{equation}
{\rm Prob}^{(1)}( 1 \rightsquigarrow 2) \geq  1 -\exp \left[ - n K \Lambda^{-aj} r_{12}^2 \right]
\label{eq4}
\end{equation}
where $K = 2 \pi/3 -\sqrt{3}/4$.  

Now, clearly, $R_0 \Lambda^j < r_{12} \Lambda$. Hence we have $\Lambda^{-a j} > r_{12}^{-a} \Lambda^{-a}/R_{0}^{-a}$.  Putting this in Eq. \ref{eq4}, we get 
\begin{equation}
{\rm Prob}^{(1)}( 1 \rightsquigarrow 2) \geq  1 - \exp\left[ -n K \Lambda^{-a} r_{12} ^{-a +2} \right]
\label{eq5}
\end{equation}
where some constant factors have been absorbed into $K$. For large $r_{12}$, it is easy to see that this varies as $r_{12}^{2 -a}$. Hence the two-point correlation function is bounded from below by a power-law.

We can extend this calculation, and write the two-point correlation function as an expansion
\begin{equation}
{\rm Prob}( 1 \rightsquigarrow 2) = \sum_{n=1}^{\infty} {\rm Prob}^{(n)}(1\rightsquigarrow 2) 
\label{eq6}
\end{equation}
where ${\rm Prob}^{(n)}(1\rightsquigarrow 2)$ is the probability that the path of overlapping discs connecting points $P_1$ and $P_2$ requires $n$ discs. The term $n=2$ corresponds to a more complicated integral over two overlapping discs. But it easy to see that for large $r_{12}$, this  also decays as 
$r_{12}^{-a +2}$.  Assuming that similar behavior holds for higher order terms as well, we expect that  for all non-zero densities $n$, the two-point correlation function decays as a power law even for arbitrarily low densities of discs. 

We note that this is consistent with the result that for continuum percolation in $d$ dimensions, the diameter of the connected component containing the origin say $\langle D \rangle$ is divergent even for arbitrarily small number densities when $\langle R^{d +1}\rangle$ is divergent \cite{gouere}. Here $R$ denote the radii variable.  In our case $\langle D \rangle=\int r_{12}\dfrac{d Prob(r_{12})}{d r_{12}}dr_{12} \sim  \int r_{12}^{2 - a} dr_{12}$ (where $P_1$ is the origin) is divergent when $a\leq 3$, consistent with the above. 

The power-law decay of the two-point function is the result of the fact that for any distance $r$, we have discs of radii of the order of $r$. However for large values of $r$, we can imagine that there would also be a contribution from a large number of overlapping discs of radii much smaller than $r$ connecting the two points separated by the distance $r$, which as in the usual percolation problem decays exponentially with distance. Therefore it is reasonable to write the two point function in our problem as a sum of two parts; the first part say $G_{sr}(r)$ due to the \textquoteleft short range\textquoteright\;  connections which has an exponential decay with distance for large $r$ and the second one say $G_{lr}(r)$ due  to the \textquoteleft long range\textquoteright\;  connections which has a power law decay with distance. Therefore,
\begin{equation}
 G(r) = G_{sr}(r) + G_{lr}(r)
\label{eq9}
\end{equation}
where
\begin{equation}
 G_{lr}(r) \sim D(A)/r^{a-2} + higher\;order\;terms
\label{eq10}
\end{equation}
where $D(A)$ is assumed  to go to a non-zero constant as $A$ goes to its critical value and its dependence on $A$ is a slowly varying one. %Here $t$ is equal to $a-2$  but may also include an anomalous dimension exponent.
%While Eq.(\ref{eq9}) defines the correlation length in the problem, %[In systems with a power-law decay for the two-point function, the correlation length can be defined through the second moment of the two-point function or in cases where even that doesn't exist by, $\int G(r) dr \sim 1/\xi^{t}$;  see for eg. \cite{stell1}],
%the important question is that whether the critical behavior of the system is affected by the \textquoteleft long range\textquoteright\; connections.

The power-law distribution of the radii, makes this system similar to a long range interaction problem in statistical physics in the sense that given two points in the plane, a direct connection by a single disc overlapping both the points is possible. In fact similar behavior for the two-point function exists whenever we have long range interactions in a system, such as in Ising model with long range potentials or fluid with long range interactions \cite{sak,kayser}. In such systems, the two-point function shows a power-law decay just as in our problem \cite{iagolnitzer}. The effect of such long range potentials on the critical exponents have been studied earlier \cite{stell1, stell2, fisher, aizenman, sak,dantchev} with the general conclusion that  the long range part of the interaction do influence  the critical behavior of the system \cite{stell3}. More precisely, if we have an attractive pair potential in $d$ dimensions of the form $-\phi(r) \sim \dfrac{1}{r^{d + \sigma}}$ where $\sigma > 
0$, then critical exponents take their short-range values for all $\sigma \geq 2 - \eta_{sr}$ where $\eta_{sr}$ is the anomalous dimension (For the \textquoteleft short range\textquoteright\; problem in 2D, at criticality, the two point function decays with distance as $1/r^{\eta_{sr}}$). For $\sigma < 2-\eta_{sr}$, two kinds of behavior exist. For $0 < \sigma \leq d/2 $, the exponents take their mean-field values and for $ d/2 < \sigma < 2 - \eta_{sr}$, the exponents depend on the value of $\sigma$ (See \cite{aizenman} and references therein). So $\sigma = 2 - \eta_{sr}$ is the dividing line between the region dominated by short range interactions and the region dominated by long-range interactions.

Though there is a well established connection between the lattice percolation problem and the Ising model \cite{fortuin}, there is no similar result connecting the continuum percolation problem to any simple Hamiltonian system. However, the following simple argument provide us with a prediction about the values of the parameter $a$ for which the power-law nature of the distribution is irrelevant and the system is similar to a continuum percolation system with a bounded size distribution for the percolating units. Assuming that the strength of the long range interaction from a given point in the Ising/fluid system (which decays like $\sim \dfrac{1}{r^{2 + \sigma}}$ in 2D) is like the strength of the connectivity from the center of a given disc which is given by the distribution of the radii; in our problem, we expect the dividing line between the region dominated by short-range connectivity and the region dominated by long-range connectivity to be the same as that for an Ising system with long range potential 
of the form $-\phi(r) \sim \dfrac{1}{r^{a+1}}$ where $a>2$. Then the results for the long-range Ising system discussed in the last paragraph should carry over with $\sigma = a - 1$. So for our problem, a deviation from the standard critical behavior is expected when $a<3-\eta_{sr}$  and the critical exponents will take their short-range values for $a>3-\eta_{sr}$. For 2D percolation, $\eta_{sr} = 5/24$ \cite{nienhuis}. Also mean-field behavior is expected when $a\leq2$. However for this range of $a$, the entire plane is covered for all non-zero number densities and hence there is no phase transition.
%Though there is a well established connection between the lattice percolation problem and the Ising model \cite{fortuin}, there is no similar result connecting the continuum percolation problem to any simple Hamiltonian system. However, simulation studies  suggest that the continuum percolation problem (where percolating units are bounded in size) and the lattice percolation problem belong to the same universality class \cite{vicsek}. In our problem, the behavior is similar to that of an Ising system with long range potential of the form $-\phi(r) \sim \dfrac{1}{r^{a-1}}$ where $a>2$. Then the results from \cite{stell3} shows that a deviation from the standard critical behavior is expected when $a<3-\eta_{sr}$  and the critical exponents will take their short-range values for $a>3-\eta_{sr}$. {\color{red} For 2D percolation, $\eta_{sr} = 5/24$ \cite{nienhuis}}. Also mean-field behavior is expected when $a\leq2$. However for this range of $a$, the entire plane is covered for all non-zero number densities and 

In the next two sections,  we investigate for the dependence of exponents on  the power-law tail of the distribution of the radii of the discs. First we develop an approximate RG method. Then we carry out  simulation studies  which show that the correlation length exponent $\nu$ takes its short range value for $a>3 - \eta_{sr}$, while it depends upon $a$ for $a<3 - \eta_{sr}$. %This shows that the 'long range' part of the connectivity does affect the critical behavior of continuum percolation of discs with a distribution of radii having a fat-tailed distribution. %However, we note that an unambiguous definition of critical exponents them self  and the scaling relations between them will depend upon the value of the parameter $a$. For eg, while we can define $\nu$ for all $a>3$,we can define $\beta$ only for $a>4$ since for all $a\leq4$, $\langle |C|\rangle$ is divergent CITE{}.
 
\section{Critical behavior near the percolation threshold}
\label{sec5}

In this section, we propose an approximate RG method to analyze the behavior of continuum percolation models near the percolation threshold, when the percolating units have a distribution of sizes. We assume that we can replace discs of one size having a number density $n$ with discs of another size and  number density $n'$, provided the correlation length remains the same. Application of a similar idea in  disc percolation  problem with only two sizes of discs may be found in \cite{balram}.

We will illustrate the method by considering a problem in which the radii of discs take only two possible values, say $R_{1}$ and $R_{2}$. Let their areal densities be $A_{1}$ and $A_{2}$ respectively, and assume that both $A_{1}$ and $A_{2}$ are below $A^{*}$, the critical threshold for the percolation problem with only single sized discs present ( $A^{*} \approx 1.128085$ \cite{quintanilla2}). Also let $\xi_{1}$ represent the correlation length when only discs of size $R_{1}$ are present in the system and $\xi_{2}$ represent that when only discs of size $R_{2}$ are present. Invariance of the two point function under length rescaling requires that the expression for the correlation length $\xi$ is of the form $\xi = Rg(A)$, where the function $g(A)$ determines how the correlation length depends on the areal density $A$ and is independent of the radius $R$. Let $\tilde{A}_{2}$ is the areal density of the discs of size $R_{2}$ which will give the same correlation length as the discs of size $R_{1}$. i.e,
\begin{equation}
\xi_{1}\left(A_{1}\right) = \xi_{2}\left(\tilde{A}_{2}\right)
\label{eq11}
\end{equation}
or
\begin{equation}
R_{1}g\left(A_{1}\right) = R_{2}g\left(\tilde{A}_{2}\right)
\label{eq12}
\end{equation}
Given the form of the function $g(A)$, we can invert the above equation to find $\tilde{A}_{2}$. Formally,
\begin{equation}
\tilde{A}_{2} = g^{-1}\left(\frac{R_{1}}{R_{2}}g\left(A_{1}\right)\right)
\label{eq13}
\end{equation}
So the problem is reduced to one in which only discs of size $R_{2}$ are present, whose net areal density is now given by,
\begin{equation}
 A_{2}^{\prime} = \tilde{A}_{2} + A_{2}
\label{eq14}
\end{equation}
System percolates  when ${A}_{2}^{\prime} = A^{*}$.
Now, when areal density $A$ is close to $A^{*}$, we have
\begin{equation}
 g(A)= C \left(A^{*} - A\right)^{-\nu}
\label{eq15}
\end{equation}
where $C$ is some constant independent of $A$ and $\nu$ is the correlation-length exponent in the usual percolation problem. Using this in Eq. \ref{eq13}, we get
\begin{equation}
\tilde{A}_{2} = A^{*} - \left(A^{*} - A_{1}\right)\left(R_{2}/R_{1}\right)^{1/\nu}
\label{eq16}
\end{equation}
Therefore, for a given value of $A_{1}<A^{*}$, the areal density of discs of radius $R_{2}$,  so that the system becomes critical is given by,
\begin{align}
 \nonumber A_{2} &=A^{*} - \tilde{A}_{2} \\ 
 &=\left(A^{*} - A_{1}\right)\left(R_{2}/R_{1}\right)^{1/\nu}
\label{eq17}
\end{align}
So the total areal density at the percolation threshold is, 
\begin{align}
 \nonumber A_{1} + A_{2} &= A_{1} + \left(A^{*} - A_{1}\right)\left(R_{2}/R_{1}\right)^{1/\nu}\\
\nonumber &= A_{1}(1-x) + A^{*}x
\end{align}
where $x = \left(R_{2}/R_{1}\right)^{1/\nu}$. Without loss of generality we may assume $R_{2}>R_{1}$. Then $x>1$ and we can see from the above expression that the percolation threshold $A_{1} + A_{2}> A^{*}$, a result well known from both theoretical studies \cite{meester2} and simulation studies \cite{quintanilla2}.

Now in our problem assume that areal density of discs of type $0$ do not exceed $A^{*}$. Renormalizing discs up to type $m$ in our problem gives the equation for the effective areal density of the $m$-th type discs $A_{m}^{\prime}$ as
\begin{equation}
 A_{m}^{\prime} = A^{*} - \left(A^{*} - A_{m-1}^{\prime}\right)\Lambda^{1/\nu} + \rho_{m}
\label{eq18}
\end{equation}
where $m\geq1$, ${A}_{0}^{\prime} = \rho_{0}$ and $\rho_{m} = n_{0}\pi\Lambda^{(2-a)m}$ denote the areal density of discs of radius $\Lambda^{m}$. Here $n_{0}$ is the number density of discs of radius $R_{0}$ (or of type $0$), which for convenience we have set equal to unity. If we denote $A^{*} - A_{m}^{\prime}$ by $\varepsilon_{m}$ which is the distance from the criticality after $m$-th step of the renormalization, then the above expression becomes
\begin{equation}
 \varepsilon_{m} = \varepsilon_{m-1}\Lambda^{1/\nu} - \rho_{m}
\label{eq19}
\end{equation}
\begin{figure}
%\begin{center}
\includegraphics[scale=.70]{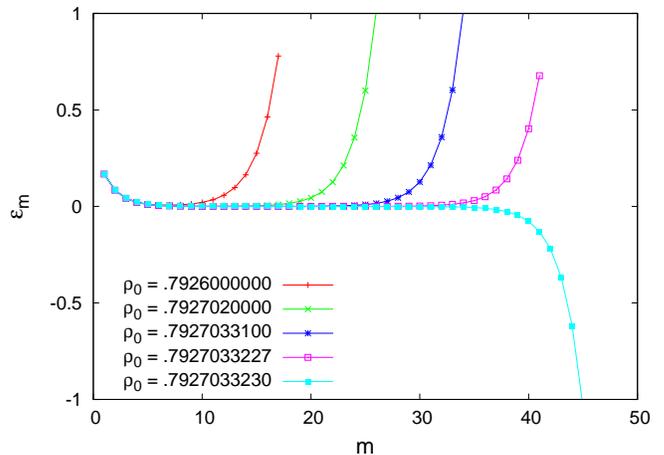} 
\caption{Variation of $\varepsilon_{m}$ with $m$ for different values of $\rho_{0}$ showing sub critical and supercritical regimes. We have used $a=3$ and $\Lambda=2$.}
\label{fig2}
%\end{center}
\end{figure}

The equation describes the flow near the critical point when we start with a value of $\rho_{0}$, the areal density of the first type of discs. Here $\varepsilon_{m}$ gives the effective distance from criticality of the $m$-th order discs in the system, in which now only $m$-th and higher order discs are present. Now for given values of the parameters $a$ and $\Lambda$, we can evaluate $\varepsilon_{m}$ in Eq. \ref{eq19} using a computer program and plot $\varepsilon_{m}$ verses $m$. Depending upon the value of $\rho_{0}$, we get three different behaviors. For value of $\rho_{0}$ below the critical value denoted by $\rho_{0}^{*}$,  $\varepsilon_{m}$ will go to $A^{*}$ asymptotically (System is sub critical) and when it is above $\rho_{0}^{*}$, $\varepsilon_{m}$ will go to $-\infty$ asymptotically (System is super critical). As $\rho_{0} \rightarrow \rho_{0}^{*}$, we get the critical behavior characterized by $\varepsilon_{m}$ tending to the RG fixed point $0$ asymptotically. Typical result using Eq \ref{eq19}
 with $\Lambda = 2$ and $a=3$ is shown in Fig \ref{fig2}.  We can see that as we tune $\rho_{0}$, the system approaches criticality, staying closer to the $\varepsilon_{m}=0$ line longer and longer. Critical behavior here can be characterized by the value of $m$ at which the curve deviates from the approach to $\varepsilon_{m} = 0$ line. To understand how the correlation length diverges as we approach criticality, we assume that we can replace the sub critical system with a system where only discs of type $m'$ is present and has a fixed areal density below $A^{*}$, where $m'$ is the value of $m$ at which  $\varepsilon_{m}$ shows a substantial increase - say  $\varepsilon_{m}$ becomes $A^{*}/2$. For continuum percolation problem with single sized discs, the correlation length $\xi = Rg(A)$, where $g(A)$ is a function with no explicit dependence on radius $R$. Therefore, correlation length in our problem, 
\begin{equation}
\xi \varpropto \Lambda^{m'}
\label{eq20}
\end{equation}

\begin{figure}[t]
%\begin{center}
\includegraphics[scale=.72]{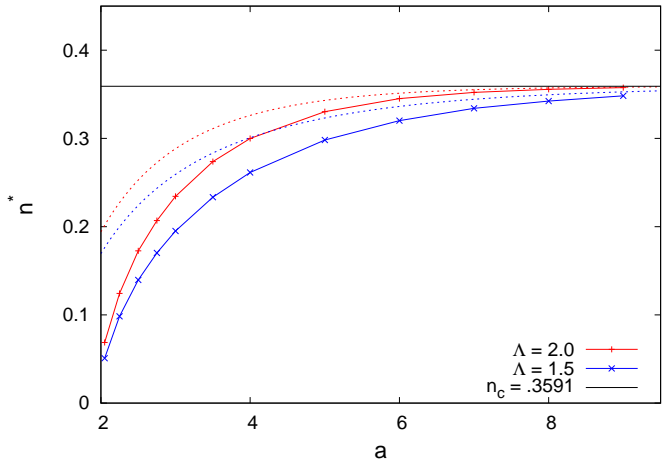} 
\caption{Variation of $n^{*}$ with $a$ for two different values of $\Lambda$. Dashed curves correspond to values given by Eq. \ref{eq30} and continuous ones correspond to those from simulation studies. The horizontal line corresponds to the threshold for the single sized discs case.}
\label{fig3}
%\end{center}
\end{figure}

We can write the recurrence relation Eq.(\ref{eq19}) in terms of the areal density $\rho_{n}$ as
\begin{equation}
\varepsilon_{m} = A^{*}\Lambda^{\frac{m}{\nu}} - \sum_{n=0}^{m}\rho_{n}\Lambda^{\left[\frac{m-n}{\nu}\right]}
\label{eq21}
\end{equation}
But $\rho_{n} = \rho_{0}\Lambda^{n\left(2-a\right)}$. Therefore,
\begin{equation}
\varepsilon_{m} =  A^{*}\Lambda^{\left[\frac{m}{\nu}\right]} - \dfrac{\rho_{0}\Lambda^{\left[\frac{m}{\nu}\right]}\left[1 - \Lambda^{m\left(2-a-1/\nu\right)}\right]}{\left[1-\Lambda^{\left(2-a-1/\nu\right)}\right]}
\label{eq22}
\end{equation}
For large values of $m$, the last term in the above equation involving $\Lambda^{m\left(2-a-1/\nu\right)}$ can be neglected. Then,
\begin{equation}
\varepsilon_{m} =  \Lambda^{\left[\frac{m}{\nu}\right]}\left[A^{*} - \frac{\rho_{0}}{1 - \Lambda^{\left(2-a-1/\nu\right)}}\right]
\label{eq23}
\end{equation}
Therefore,
\begin{equation}
\Lambda^{\left[\frac{m}{\nu}\right]} = \dfrac{\varepsilon_{m}}{\left[A^{*} - \dfrac{\rho_{0}}{1 - \Lambda^{\left(2-a-1/\nu\right)}}\right]}
\label{eq24}
\end{equation}
For a given value of $\rho_{0}\leq A^{*}$, the order $m'$ at which $\varepsilon_{m}$ is increased substantially, say to a value $A^{*}/2$ is given by
\begin{equation}
\begin{split}
m' &= [\log_{\Lambda}\left(A^{*}/2\right) - \log_{\Lambda}\left(\rho_{0}^{*} - \rho_{0}\right) \\
&\quad + \log_{\Lambda}\left(1 - \Lambda^{\left(2-a-1/\nu\right)}\right)]\nu 
\label{eq25}
\end{split}
\end{equation}
So for $\rho_{0}$ close to $\rho_{0}^{*}$ and large values of $a$,
\begin{equation}
 m' \sim \log_{\Lambda}\left(\rho_{0}^{*} - \rho_{0}\right)^{-\nu}.
\label{eq26}
\end{equation}
so that 
\begin{equation}
\xi \varpropto \left(\rho_{0}^{*} - \rho_{0}\right)^{-\nu}
\label{eq27}
\end{equation}
Thus we find that the correlation length exponent $\nu$ is independent of the parameters $a$ and $\Lambda$ of the distribution. From Eq. \ref{eq24}, we can also obtain the percolation threshold $\rho_{0}^{*}$ as a function of the parameters $a$ and $\Lambda$. In Eq. \ref{eq24} left hand side is positive definite. So for values of $\rho_{0}$ for which $\frac{\rho_{0}}{1 - \Lambda^{\left(2-a-1/\nu\right)}} < A^{*}$, we will have $\varepsilon_{m}>0$ for large values of $m$. Similarly for values of $\rho_{0}$ for which $\frac{\rho_{0}}{1 - \Lambda^{\left(2-a-1/\nu\right)}} > A^{*}$, we will have $\varepsilon_{m}<0  $ for large values of $m$. Hence the critical areal density $\rho_{0}^{*}$ must be given by
\begin{equation}
 \rho_{0}^{*} = A^{*}\left[1 - \Lambda^{\left(2-a-1/\nu\right)}\right]
\label{eq29}
\end{equation}
Or in terms of the total number density, the percolation threshold $n^{*}$ is given by,
\begin{equation}
n^{*} =     n_c\left(1 - \Lambda^{\left(2-a-1/\nu\right)}\right)/\left(1 - \Lambda^{-a}\right)   
\label{eq30}
\end{equation}
where $n_c = A^{*}/\pi$, the critical number density for percolation with single sized discs of unit radius. Note that this approximate result does not give the correct limit, $n^{*} \rightarrow 0$ as $a \rightarrow 2$. The RG scheme depends on the approximation that the effect of size $R_1$ of areal density $A_1$ is the same as that of discs of radius $R_2$ of density $A_2$, as in Eq. \ref{eq11}. This is apparently good only for $a > 3 - \eta_{sr}$.  Fig. \ref{fig3} shows the variation of the critical threshold with $a$ for two different values of $\Lambda$ using Eq. \ref{eq30} along with simulation results (See section \ref{sec6} for details of simulation studies). We see that a reasonable agreement is obtained between the two for higher values of $a$.  Also, as one would expect, for large values of $a$, $n^{*}$ tends to  $n_c$.

From Eq. \ref{eq30}, we can obtain the asymptotic behavior of the critical number density $n^*$ as $\Lambda \rightarrow 1$. This is useful since it corresponds to the threshold for a continuous distribution of radii with a power-law tail and we no more have to consider the additional discretization parameter $\Lambda$. It is easy to see that in the limit $\Lambda \rightarrow 1$, Eq. \ref{eq30} becomes

\begin{equation}
 n^{*}_{\Lambda \rightarrow 1} = n_c\left(1 - \dfrac{5}{4 a}\right) 
\label{eq30_2}
\end{equation}
where we have used the value $\nu = 4/3$. Thus we expect that a log-log plot of $(n_c - n^*_{\Lambda \rightarrow 1})$ against $a$ will be a straight line with slope $-1$ and y-intercept $\log(5 n_c/4) \approx -0.35$ for large values of $a$.  A comparison  with the thresholds  obtained from simulation studies show that Eq. \ref{eq30_2} indeed predicts the asymptotic behavior correctly (See Fig. \ref{threshold}).

\section{Simulation Results}
\label{sec6}

We determine the exponent $\nu$ and the percolation threshold $n^{*}$ by simulating the continuum percolation system in 2D, with  discs having a power law distribution for their radii.  We consider two cases for the distribution of the radii variable. To explicitly compare the prediction of the approximate RG scheme for the percolation threshold given in section \ref{sec5}, we  use a discrete distribution for the radii variable, with discretization factor $\Lambda$ as in section \ref{sec2}. The results for the thresholds thus obtained is shown in Fig. \ref{fig3}. To determine the correlation length exponent $\nu$, we  consider the radii distribution in the limiting case $\Lambda \rightarrow 1$, so that we do not have to consider the additional parameter $\Lambda$. In this case, given a disc, the probability that it has a radius between $R$ and $R+dR$ is equal to  $a R^{-(a+1)}$ where $a>2$. We  also obtain the percolation threshold with this continuous distribution for the radii and compare it with the predicted asymptotic behavior in Eq. \ref{eq30_2}. The minimum radius is assumed to be unity. 

For $a\leq2$ the entire plane is covered for arbitrarily low densities of the discs. We use cyclic boundary conditions and consider the system as percolating whenever it has a path through the discs from the left to the right boundary. We drop discs one at a time on to a region of a plane of size $L\times L$, each time checking whether the system has formed a spanning cluster or not. Thus number density is increased in steps of $1/L^2$. So after dropping the $n-th$ disc, the number density is $n/L^2$. Now associated with 
each number density we have a counter say $f_n$ which is initialized to $0$ in the beginning. If the system is found to span after dropping the $n'$-th disc, then all counters for $n\geq n'$ is incremented by one. After a spanning cluster is formed, we stop. By this way we can determine the spanning probability $\varPi(n,L) = f_n/N$ where $N$ is the number of realizations sampled. The number of realizations sampled varies from a maximum of $2.75\times10^7 $ for $a = 2.05$ and $L = 90$ to a minimum of $4000$ for $a = 10.0$ and $L = 1020$ [For obtaining the results for the threshold in Fig. \ref{fig3}, the number of realizations sampled is $20000$ for all values of $a$ and $\Lambda$]. This method of dropping basic percolating units one by one until the spanning cluster is formed has been used before \cite{li} in the context of stick percolation which was based on the algorithm developed in \cite{newman}, and allows us to study relatively large system sizes with large number of realizations  
within reasonable time. 

\begin{figure}
%\begin{center}
\includegraphics[scale=0.70]{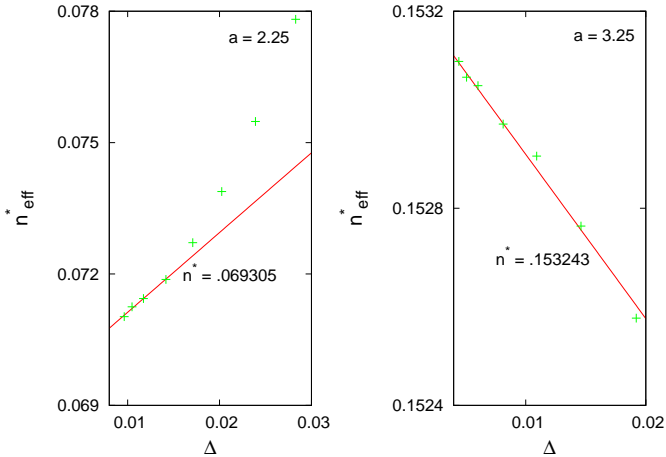} 
\caption{Plot of effective percolation threshold $n^*_{eff}$ against $\Delta$ for $a = 2.25$ and $a = 3.25$. The best straight line fit is obtained with the last four data points.}
\label{fig6}
%\end{center}
\end{figure}

\begin{figure}
%\begin{center}
\includegraphics[scale=0.70]{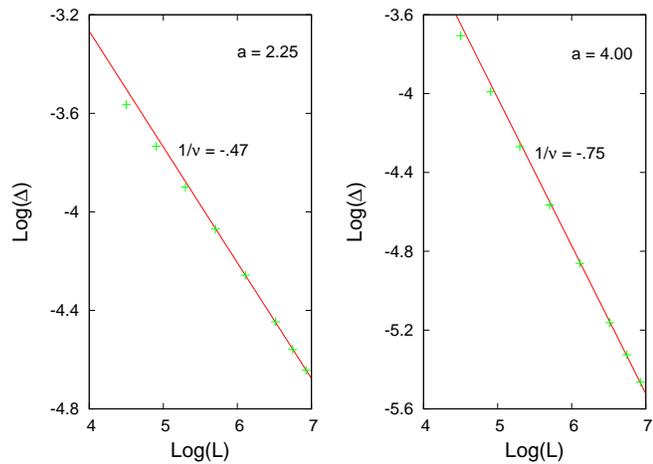} 
\caption{Log-Log plot of $\Delta$ Vs $L$ for $a = 2.25$ and $a = 4.0$ along with lines of slope $-.47$ and $-.75$.}
\label{fig7}
%\end{center}
\end{figure}

The probability that there is at least a single disc which span the system of size $L$ at number density $n$ is $1 - \exp^{\left(-n 2^{a}/L^{a-2}\right)}$. It is easy to see that to leading order in $n$, this \textquoteleft long range\textquoteright\,  part of the spanning probability $\varPi(n,L)_{lr}$ is $\dfrac{n 2^{a}}{L^{a-2}}$. So one can write a scaling form for the spanning probability,
\begin{equation}
\varPi(n,L) = \varPi(n,L)_{lr} + (1 - \varPi(n,L)_{lr}) \phi((n^{*} - n)L^{1/\nu})
\end{equation}
Therefore we can define the \textquoteleft short range\textquoteright\, part of the spanning probability 
 $\varPi^{\prime}(n,L) = (\varPi(n,L)-\varPi(n,L)_{lr})/(1-\varPi(n,L)_{lr})$
where the leading long range part is subtracted out. Therefore, we have 
\begin{equation}
\varPi^{\prime}(n,L) = \phi((n^{*} - n)L^{1/\nu})
\label{eq31}
\end{equation}
and the scaling relations, (See for eg. \cite{stauffer})
\begin{equation}
\Delta(L) \varpropto L^{-1/\nu}
\label{eq32}
\end{equation}
\begin{equation}
n^*_{eff}(L) - n^{*} \varpropto \Delta
\label{eq33}
\end{equation}
where $n^*_{eff}(L)$ is a suitable defined effective percolation threshold for the system of size $L$, and  $\Delta$ is the width of the percolation transition obtained from the spanning probability curves $\varPi^{\prime}(n,L)$. Note that Eqs. \ref{eq32} and \ref{eq33} are applicable with any consistent definition of the effective percolation threshold and width $\Delta$ \cite{stauffer}. A good way to obtain $n^*_{eff}$ and $\Delta$ is to fit the sigmoidal shaped curves of the spanning probability  $\varPi^{\prime}(n,L)$ with the function $1/2[1 + erf[(n-n^*_{eff}(L))/\Delta(L)]]$ (see \cite{rintoul2}), which defines the effective percolation threshold $n^*_{eff}$ as the number density at which the spanning probability is $1/2$. We determined $n^*_{eff}$ and $\Delta$ for each value of $a$ and $L$ and determined  $1/\nu$ and $n^*$ for different values of $a$ using Eqs. \ref{eq32} and \ref{eq33} respectively. Typical examples are shown in fig. \ref{fig6} and Fig. \ref{fig7}.  

\begin{figure}
%\begin{center}
\includegraphics[scale=.70]{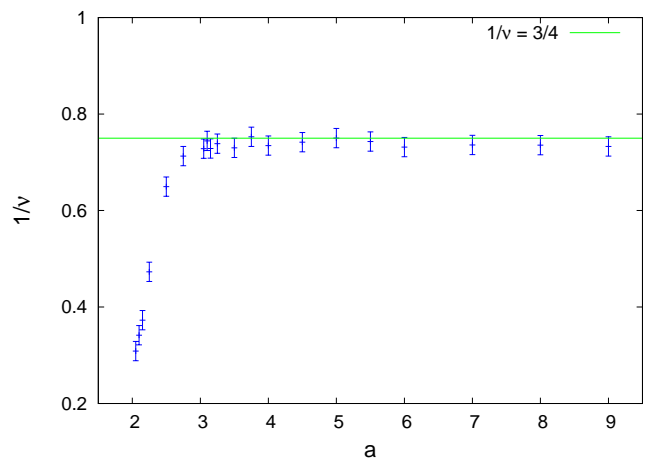} 
\caption{Variation of $1/{\nu}$ with $a$. The horizontal line corresponds to the standard 2D percolation value $1/{\nu} = 3/4$.}
\label{1_nu}
%\end{center}
\end{figure}

\begin{figure}
%\begin{center}
\includegraphics[scale=.70]{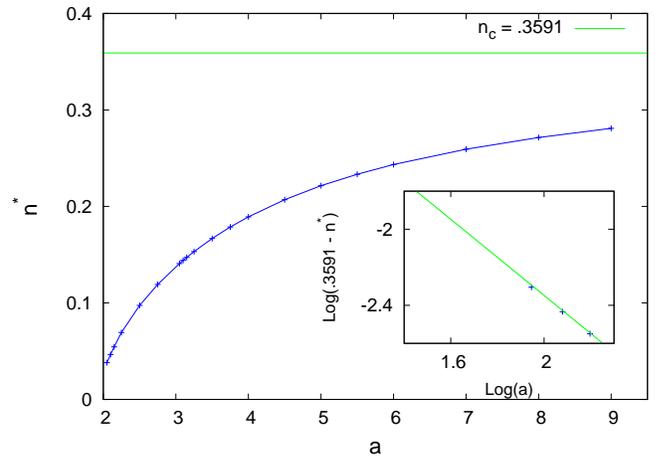} 
\caption{Variation of percolation threshold $n^*$ with $a$. The horizontal line corresponds to the threshold for the single sized discs case. (Inset) Asymptotic approach of $n^*$ to the single sized discs value $n_c = .3591$  along with a straight line of slope $-1$ and y-intercept $-0.35$ (See Eq. \ref{eq30_2}).}
\label{threshold}
%\end{center}
\end{figure}

At first, we determined the percolation threshold and the exponent for a system of single sized discs of unit radius. We obtained $n^* = .3589(\pm.0001)$ (or areal density  $\approx 1.12752$) and $1/\nu = .758(\pm.018) $ in very good agreement with the known value for the threshold \cite{quintanilla2} and the conjectured value of $1/\nu = 3/4$ for the exponent. Values of $1/\nu$ obtained for various values of $a$ are shown in fig.\ref{1_nu}. We scan the low $a$ regime  more closely for any variation from the standard answer. We can see that the estimates for $1/\nu$ are very much in line with the standard percolation value for $a>3 - \eta_{sr}$ while it varies with $a$ for $a<3 - \eta_{sr}$.  Fig. \ref{threshold} shows the variation of the percolation threshold $n^*$ with $a$. As expected, with increasing $a$, the percolation threshold increases and tends to the single sized disc value as $a\rightarrow \infty$, and as $a\rightarrow 2$, the threshold tends to zero. The data also shows that $n^*$ converges to 
the threshold for the single sized disc value as $1/a$ as predicted by Eq. \ref{eq30_2}. Values of the threshold for some values of $a$ are given in Table \ref{table1}.

\begin{table}[ht]
 \centering
\begin{tabular}{| c |  r | c | r|}
\hline
$a$ & \multicolumn{1}{c |}{$n^*$} & $\eta^* = n^*\pi a/(a-2)$ & $\phi^* = 1 - \exp^{-\eta^*}$
\\[0.5ex]
\hline
2.05	&	0.0380(6)    &     4.90(7)		&        0.993(1)\\
2.25	&	0.0693(1)     &    1.959(3)		&	0.8591(5)\\
2.50	&	0.09745(11)      &   1.5307(17)		&	0.7836(4)\\
3.50	&	0.16679(8)      &     1.2226(6)	&	0.70555(17)\\
4.00	&	0.18916(3)      &     1.1885(2)	&	0.69543(6)\\
5.00	&	0.22149(8)      &     1.1597(4)	&	0.68643(13)\\
6.00	&	0.24340(5)      &     1.1470(2)	&	0.68241(8)\\
7.00	&	0.2593(2)     &   1.1406(7)		&	0.6804(2)\\
8.00	&	0.27140(7)      &    1.1368(3)  	 &       0.67917(9)\\
9.00	&	0.28098(9)      &    1.1349(4)	&	0.67856(12)\\[1ex]
\hline
\end{tabular}
\caption{Percolation threshold $n^*$ for a few values of $a$ along with corresponding critical areal density $\eta^*$ and the critical covered area fraction $\phi^*$.}
\label{table1}  %label of the table.

\end{table}

\begin{figure}
%\begin{center}
\includegraphics[scale=0.70]{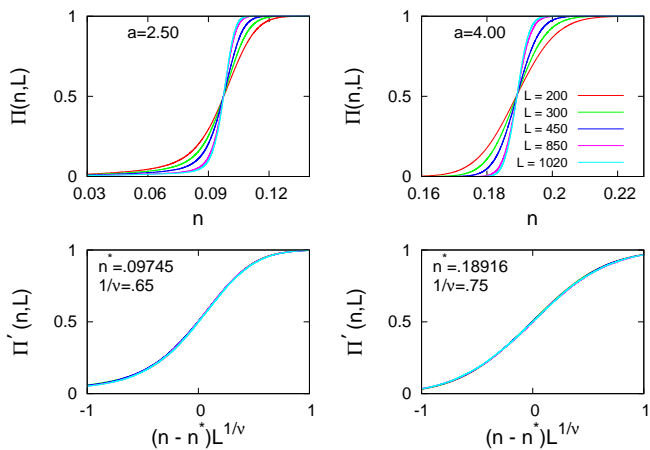} 
\caption{Variation of $\varPi(n,L)$ with $n$ (top row) and the scaling collapse (bottom row) for $a = 2.50$ (on left) and $a = 4.00$ (on right).}
\label{collapse}
%\end{center}
\end{figure}

Finally as a check, we plot the spanning probability $\varPi^{\prime}(n,L)$ (see Eq. \ref{eq31}) against  $(n-n^{*})L^{1/\nu}$ to be sure that a good scaling collapse is obtained.  We show two such plots for $a=2.50$ and $a=4$ in fig. \ref{collapse}. We can see that a very good collapse is obtained. Similar good collapse is obtained for other values of $a$ as well.

\section{concluding remarks}
\label{sec7}
In this paper, we discuss the effect of a power-law distribution of the radii on the  critical behavior of a disc percolation system. If the distribution of radii is bounded, then one would expect the critical exponents to be unchanged and would be the same as that for standard percolation.  However, if the distribution of radii  has a power-law tail, we show that this strongly influence the  nature of the phase transition. The whole of the low-density non-percolating phase has power-law decay of correlations in contrast to the exponential decay for the standard percolation and this occurs for any value of the power $a$, howsoever large. The critical exponents  depend on the value of $a$ for $a < 3 - \eta_{sr} $ and take their short-range values for $a > 3  - \eta_{sr}$. We also propose an approximate RG scheme to analyse such systems. Using this, we compute the correlation-length exponent and the percolation threshold. The approximate RG scheme is good only for $a > 3  - \eta_{sr}$.  Monte-Carlo simulation results for the percolation thresholds and the correlation-length exponent are presented. 

We can easily extend the discussion to higher dimensions, or other shapes of objects. It is  easy to see that the power law correlations will exist in corresponding problems in higher dimensions as well. 
%For eg. for a continuum percolation problem in 3D with overlapping spheres having a power-law tail for their radii distribution, the two point function will decay as $r_{12}^{3-a}$ where $a>3$.

\begin{acknowledgments}
The author would like to thank Deepak Dhar for suggesting the problem and for  valuable suggestions and comments. The author would also like to thank Robert Ziff and anonymous referees for their useful comments and suggestions on  the manuscript. The computations reported here were performed on the computational resources of  the Department of Theoretical Physics, TIFR.
\end{acknowledgments}

\end{document}